\documentstyle[12pt,mymoriond,psfig]{article}
\begin{document}
\heading{ISO MID-INFRARED OBSERVATIONS 
\\ OF ABELL 370\footnote{ISO 
is an ESA project with instruments funded by ESA Member
States (especially the PI countries: France, Germany, the Netherlands and
the United Kingdom) and with participation of ISAS and NASA}
}

\author{
L.~Metcalfe  $^{\Diamond}$, 
B.~Altieri   $^{\Diamond}$, 
H.~Aussel    $^{\ddagger}$,
A.~Biviano   $^{\Diamond,\sharp}$, \\
B.~Mc Breen  $^{\clubsuit}$, 
M.~Delaney   $^{\clubsuit}$, 
D.~Elbaz     $^{\ddagger}$,
L.~Hanlon    $^{\clubsuit}$,
M.~Kessler   $^{\Diamond}$, \\
K.~Leech     $^{\Diamond}$, 
K.~Okumura   $^{\Diamond}$,
B.~Schulz    $^{\Diamond}$,
J.-L.~Starck $^{\ddagger}$,
L.~Vigroux   $^{\ddagger}$
}
{$^{\Diamond}$ ISO Science Operations Centre, Astroph. Division, Space Science
Dept. of ESA, Villafranca, Spain.}
{$^{\ddagger}$ Service d'Astrophysique/DAPNIA/DSM, CEA-Saclay, France.}
{$^{\clubsuit}$ Department of Experimental Physics, University College, Dublin, 
Ireland.}
{$^{\sharp}$ Istituto Te.S.R.E., CNR, Bologna, Italy.}

\begin{moriondabstract}
We report on the mid-IR imaging at 7 and 10 $\mu$m
of the galaxy cluster Abell~370, obtained with 
the ISOCAM instrument onboard ESA's Infrared Space Observatory (ISO),
as part of an ongoing program to image gravitational arcs and arclets in
distant clusters.

\noindent We have recorded initial 
detections of the A0 giant arc in both bandpasses,
the two central dominant galaxies at 7 $\mu$m, and several other cluster 
members and field galaxies.

\noindent These preliminary results are 
indicative of the potential output of our
program when more extensive observational data become available.

\end{moriondabstract}

\section{Introduction}
The cluster Abell~370 is a distant ($<z>=0.374$~\cite{M88}),
rich, massive cluster, with two central giant galaxies dominating its
optical image. The X-ray luminosity of the cluster is\footnote{We assume 
a Hubble constant
$H_0=75$~km~s$^{-1}$~Mpc$^{-1}$ and a deceleration parameter $q_0=1/2$
throughout this paper, giving a luminosity-distance to the cluster of 1613~Mpc.}
$3.7 \times 10^{44}$ erg~s$^{-1}$ \cite{Henry}, which is as expected 
from the high value of the velocity dispersion of this cluster, 
1340~km~s$^{-1}$\cite{M88}.

A370 was shown by \cite{Bautz}, \cite{BO84} and \cite{M88} 
to contain an anomalous fraction of blue objects, as compared to nearby 
clusters (50~\% of the cluster members in the central region show evidence for
star-formation activity, as compared to a mere 3~\% in Coma).
Based on optical and near-IR photometry, it has been shown 
(\cite{Aragon}, \cite{Stanford}, \cite{McLean}) that
most galaxies in the cluster are reasonably well
fitted by passive evolution models, while the existence
of galaxies in a post-starburst phase is controversial (\cite{Aragon} vs.
\cite{Stanford}). The spectral energy distribution of the cluster galaxies 
must be known over a wider wavelength range before their nature can be firmly
established, and in this context ISO observations are very important.

{\psfig{file=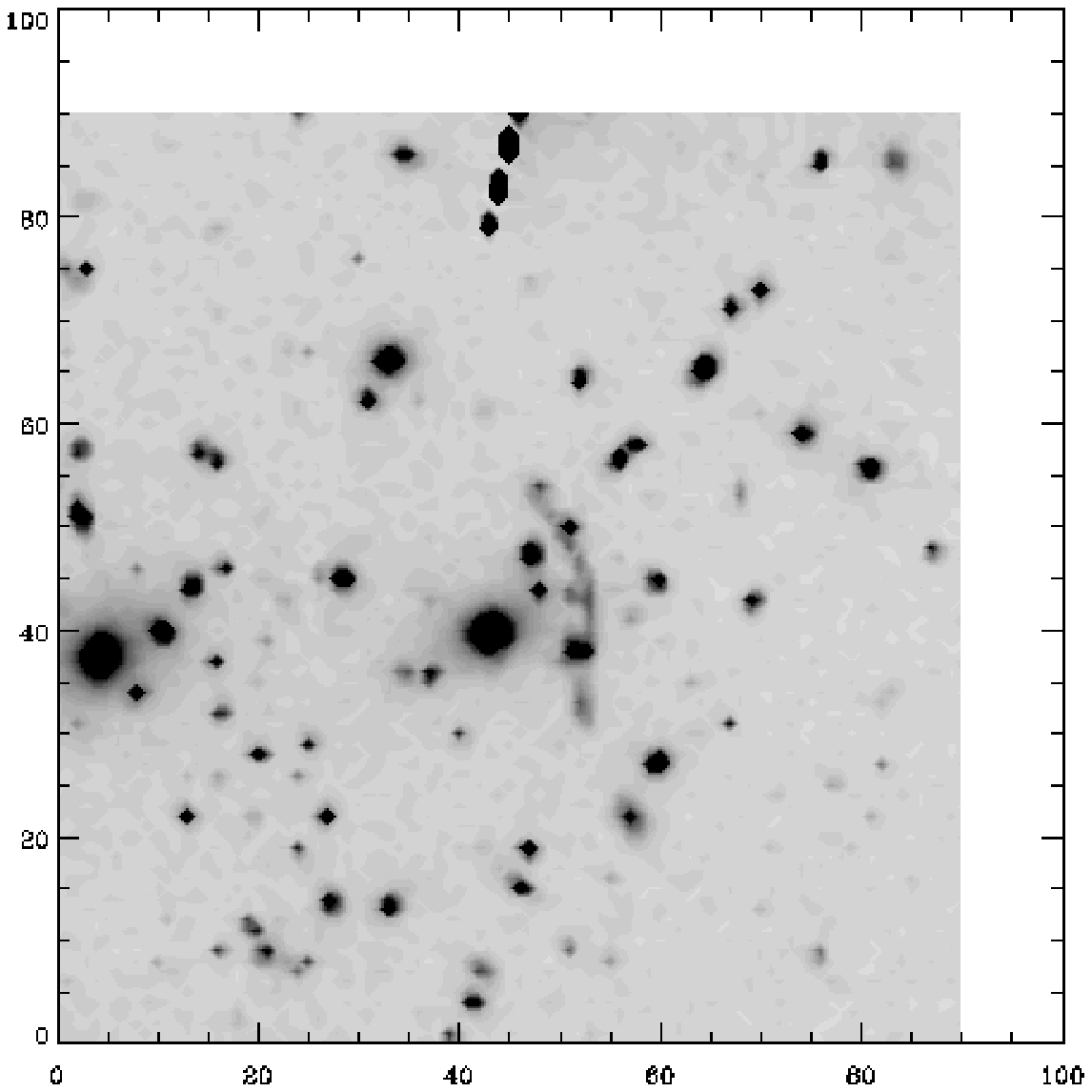,width=15.cm,angle=0}}
{\bf Fig.1.} I-band optical image of A370 (from \cite{Kneib}); the original
image has been transformed to match the ISOCAM image field-of-view. The
coordinates are arcsec. The strange feature in the upper part of the figure
is straylight from a very bright star off the image.
\medskip

A370 contains the first spectroscopically confirmed gravitational arc, A0
at $\mbox{z=0.724}$~\cite{Soucail}. Other arcs have since been detected
in the cluster (\cite{Kneib} and references therein).
Based on {\em Hubble Space Telescope} images, \cite{Smail} found evidence
for a faint spiral structure and an apparent bulge. 
The spiral nature of
the A0 arc was also indicated by its optical and near-IR colours and it
was detected by its CO line in emission by \cite{Casoli}.

In this paper we report on the mid-IR observations of A370
done with ISOCAM onboard ESA's {\em Infrared Space Observatory} (ISO) satellite.
A370 and another three clusters were selected for observation in the
context of an ISO guaranteed-time project aimed at generating mid-IR images
of known giant arcs and providing their mid-IR fluxes. 

In \S~2 we give a description of our observations and data-reductions;
in \S~3 we present the ISOCAM image, and compare it with an I-band image
in order to derive a colour map. We give our conclusions in \S~4.

\section{Observations and data reduction}

Abell 370 was observed on August 17, 1996 as part of an ISO program
(LMETCALF.ARCS) of observations of lensing galaxy clusters with arcs and
arclets: Abell 370, Abell 2218, Cl2244-02 and MS2137-23.
An important criterion for
target selection was to choose the clusters with the brightest arcs
in the optical, spatially extended and the farthest from the lens,
due to the limited spatial resolution of ISO.

The observation was performed by rastering in microscanning mode
yielding some of the deepest ISOCAM images of a galaxy cluster.
The pixel field-of-view was 3$^{\prime\prime}$ and the microscanning steps were
of 7$^{\prime\prime}$ in both directions of the 5$\times$5 raster.
The cluster was imaged in the 2 filters, LW2 [5-8.5 $\mu$m]
and LW7 [8.5-11 $\mu$m], at 5 and 10 sec integration times respectively,
for total observing times of 3353 sec (LW2) and 5705 sec (LW7),
covering a field of $2 \times 2$ arcminutes.

The data reduction was done partly within the IDL based (ISO)CAM Interactive
Analysis (CIA) package\footnote{CIA is a joint development by the 
ESA Astrophysics Division and the ISOCAM
Consortium led by the ISOCAM PI, C. Cesarsky, Direction des Sciences de
la Mati\`ere, C.E.A., France.}
\cite{Ott} and partly through the usage of C++ 
based Multi-resolution Median Transform ({\em wavelet}) 
techniques \cite{Starck}.
Cross-correlation of the two methods allows us to increase the reliability of
source detections.

In Fig.~1 we show an I-band image of A370 
(kindly provided to us in electronic form by J.P.~Kneib \cite{Kneib}), which
has been shifted, rotated, and rebinned to match the ISO-CAM field-of-view. 
The field-of-view of our observations as presented here is roughly 
$1.5^\prime \times 1.5^\prime$
centered on one of the two cluster dominant galaxies (no.9 in the list 
of \cite{Butcher}, BOW9 hereafter), the one close to the giant arc A0.

One of the data reduction problems facing us in this analysis is the difficulty
of comparing photometry obtained in or near the visible region at high spatial
resolution ($<$0.5$^{\prime\prime}$)
with the diffraction limited 7 and 10 $\mu$m
photometry of ISO, where the diameter of the first Airy ring is 
$d^{\prime\prime}=0.84 \times \lambda(\mu m)$, 
so $\simeq 6^{\prime\prime}$ in the CAM LW2 filter.
In order to make this comparison and so allow the galaxy colours to be compared
with the predictions of models, we are developing the following approach.
We take the I-band image (Fig.1) and convolve it with the relevant ISO
point-spread-function and with a 
$3^{\prime\prime} \times 3^{\prime\prime}$ square pixel -- 
representative of the 3$^{\prime\prime}$ square pixels used for these 
observations. The apparent resolution of the resulting image is well matched to
the resolution obtained in our ISO rasters, which move the detector array over 
the source in steps which are multiples of
0.333 pixels (multiples of 1$^{\prime\prime}$). 

\section{Results}
An overlay of contours of the blurred I-band image onto the 
7 $\mu$m LW2 ISO image is presented in Fig.~2.
A similar data-processing for the 10~$\mu$m image is still underway.
From Fig.2, one can clearly see a general good correspondence between the 
7 $\mu$m and the I-band images.
Pending the results of planned ISO follow-up observations, we have
considered as tentative detections those sources seen at about the 5 sigma
level ($\sim 50 \mu$Jy) in the CIA-based 
ISO image and having clear  counterparts in the I-band image, and we attach 
greatest reliability to detections confirmed in the 
{\em wavelet} tool output. Only when the follow-up 
results become available will we be able to firm-up the list of tentative 
detections reported here. Nevertheless, after extensive interactive 
examination of the data sets we believe that the majority of detections 
suggested in this paper will be confirmed when the further observations are 
analysed.

The most important detected sources are: the giant arc A0,
which shows up clearly as an elongated structure, and 
the P1 image of the same gravitationally lensed galaxy. 
The two optical dominant galaxies look very prominent at 7~$\mu$m as well.
While a full analysis of the 10~$\mu$m image is underway, 
we can nevertheless already state that A0 and P1 are detected at 10~$\mu$m
as well, while the two dominant galaxies are much fainter, and almost
disappear (particularly the one closest to the arc, BOW9).

{\psfig{file=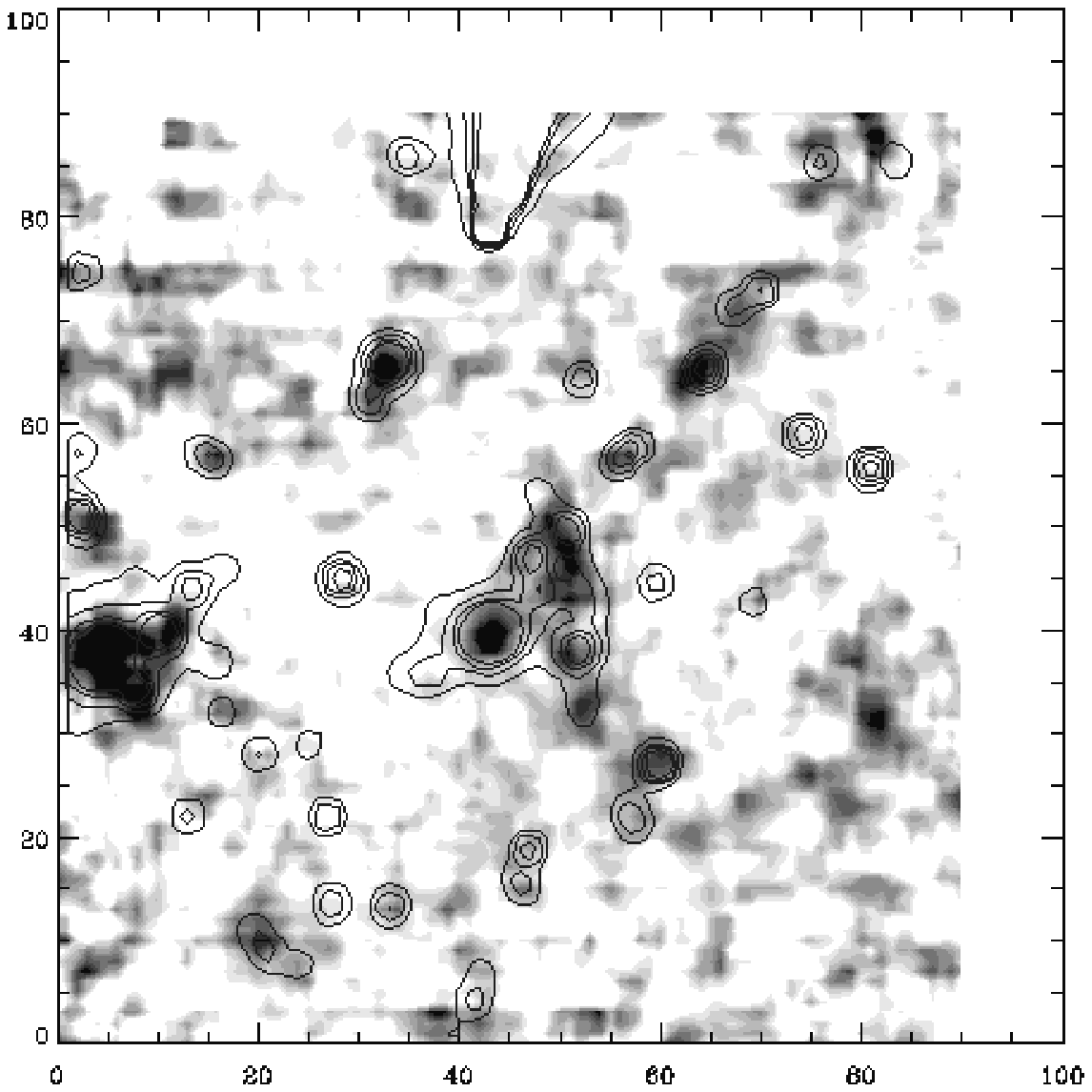,width=15.cm,angle=0}}
{\bf Fig.2} ISOCAM LW2 image (gray shading) with the blurred
I-band optical image overplotted (contours). The contours in the upper part
of the figure are an artefact due to the blurring of the straylight from
the off-image bright star.
\medskip

In Table~1 we list all the detections at 7~$\mu$m; for each detected source 
we give the coordinates in the LW2 image (Fig.~2) and 
the number of the identified galaxy in any of three catalogues
(BOW: \cite{Butcher}, MEC: \cite{MEC}, MT: \cite{McLean}). When
a single identification was not possible because of poor resolution or
confusion, while a diffuse LW2 detection was clearly present, we give the 
identification numbers of all galaxies in that region. When no galaxy was
visible in the I-band image at the location of an LW2 detection, we list
it as "IR only" in Table~1. These tentative IR detections without corresponding
optical sources are particularly interesting for future follow-up.

For the ISO image we convert to  units of Watt~m$^{-2}$
by using the conversion value from ADU/sec to mJy given in the 
"ISOCAM Observer's Manual", after a 0.8 factor correction 
as indicated by \cite{Buh}. We calibrate the I-band image via aperture 
photometry on a number of sources with optical and near-IR fluxes 
available in the literature \cite{M88}, \cite{Stanford}. 

We can then make a ratio of the ISO image over the blurred I-band image which 
yields a colour map properly reflecting the IR/optical colours of the detected
sources. The resulting colour map is presented in Fig.~3; more intense
shading corresponds to higher IR/optical flux density ratios.

From this figure it can be seen that most galaxies have a 
bluer colour than the arc (A0 and P1).
The colours corresponding to the lightest shading in the figure (0.01--0.02)
are indeed typical of normal ellipticals (according to
the models of \cite{Mazzei1}), but several galaxies are characterized by
a slight IR excess with respect to the colors of normal 10--15~Gyr old
ellipticals (including the two dominant galaxies, characterized by colours
of 0.02--0.09). The arc (A0, P1) is characterized by a color of 0.2--0.3, 
which qualifies the gravitational lensed galaxy as a spiral with intense
star formation activity, possibly in
a starburst phase (according to the models of \cite{Mazzei2} and
\cite{Granato}). The starbursting phase of the 
gravitationally lensed galaxy seems also to be confirmed by
our preliminary estimate of a high flux at 10 $\mu$m.

\pagebreak

\begin{center}
{\bf Table 1.} Detected sources in the LW2 image.
\end{center}
\begin{center}
\begin{tabular}{rrcrrc}
\hline
 x &    y  & identification & x & y & identification \\
\hline
79 &   87  & BOW167+BOW189  & 12 &   41  & MEC3                   \\  
69 &   73  & BOW130+BOW152  & 43 &   40  & BOW9, dominant galaxy  \\
52 &   66  & BOW142         &  6 &   38  & BOW10, dominant galaxy \\
12 &   65  & IR only        & 51 &   38  & MEC101                 \\
32 &   65  & BOW29          &  9 &   33  & MEC2                   \\
64 &   65  & BOW60          & 18 &   33  & BOW363                 \\
31 &   62  & BOW154         & 52 &   33  & P1 gravitational image \\
16 &   58  & BOW124         & 82 &   31  & IR only (maybe MT111)  \\
56 &   57  & MT41           & 60 &   28  & BOW119                 \\
 4 &   50  & BOW61          & 21 &   10  & BOW247+MEC25           \\
51 &   47  & Giant Arc, A0  &    &       &                        \\
\hline
\end{tabular}
\end{center}
\medskip

\section{Conclusions}
In this paper we have presented the results of observations of A370
with ISOCAM onboard ISO. We have detected the giant arc at 7 and 10 $\mu$m;
this is the first detection ever of a gravitational arc in a galaxy cluster
at these wavelengths. Moreover, we have also detected many other galaxies,
in the cluster field.

The comparison of an I-band image 
and the ISOCAM 7 $\mu$m image in the LW2 filter
has allowed us to build a colour-map which seems to indicate a
starbursting phase for the gravitationally lensed galaxy.
Most cluster galaxies however appear to have a ratio of 7 $\mu$m to
0.9 $\mu$m flux density as predicted from spectral energy
distribution models of normal ellipticals \cite{Mazzei1}.

While our data reduction methodology is still being developed and refined, 
and while a full exposition of quantitative results remains to be achieved,
the present results already demonstrate the capability of ISO in detecting 
faint mid-IR sources at large distances. With our ongoing program we hope to 
detect other gravitational arcs in clusters, and
so to constrain the evolutionary 
status of these high redshift sources. We will widen
the wavelength coverage of the spectral energy distributions
for cluster galaxies, thus providing a better leverage for constraining models 
of galaxy formation and evolution. An 
additional exciting perspective of our future 
observations is the serendipitous discovery of optically undetected "IR-only"
arcs or arclets.

\acknowledgements{We thank J.P.Kneib for providing us with a
digitized I-band image of the cluster.}

{\psfig{file=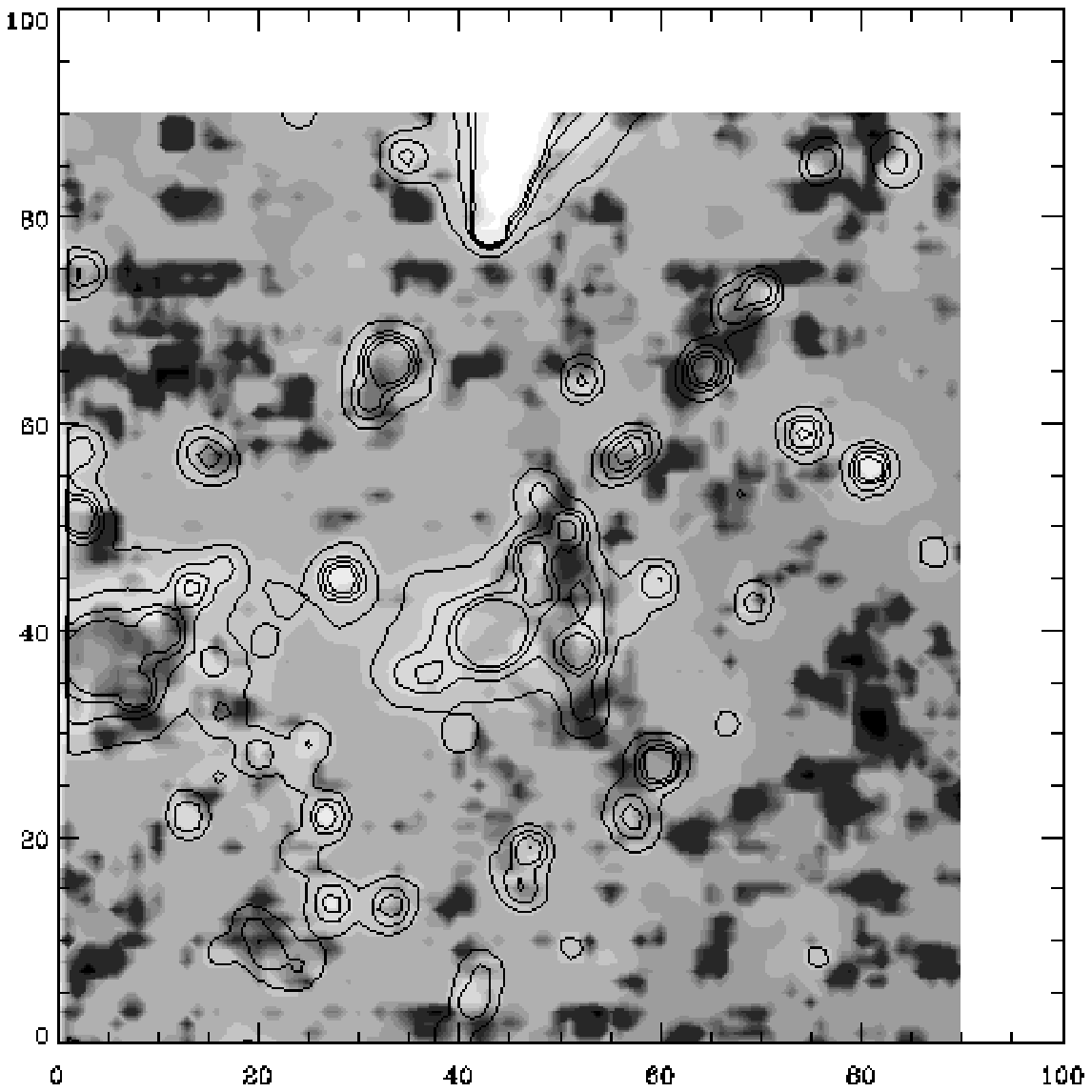,width=15.cm,angle=0}}
{\bf Fig.3} The colour map (gray shading) with the blurred
I-band optical image superposed (contours). The colour image is the
ratio of the power density in the LW2 band to the power density in the
I-band. Darker shadings correspond to higher IR to optical flux ratios.
\medskip


\end{document}